\documentclass[10pt, conference, letterpaper]{IEEEtran}
 
\IEEEoverridecommandlockouts
\usepackage{cite}
\usepackage{amsmath,amssymb,amsfonts}
\usepackage{algorithm}
\usepackage[noend]{algpseudocode}
\usepackage{graphicx}
\usepackage{textcomp}
\usepackage[table, dvipsnames]{xcolor}
\usepackage{svg}
\svgpath{{images}}
\usepackage{pdfpages}
\usepackage{tabularx}
\usepackage{subcaption}
\usepackage[export]{adjustbox}
\usepackage{tikz}
\usepackage{dblfloatfix}
\usepackage{booktabs}
\usetikzlibrary{decorations.pathreplacing}

\usepackage{caption}

\def\BibTeX{{\rm B\kern-.05em{\sc i\kern-.025em b}\kern-.08em
    T\kern-.1667em\lower.7ex\hbox{E}\kern-.125emX}}

\makeatletter
\def\BState{\State\hskip-\ALG@thistlm}
\makeatother

\renewcommand{\thesubsubsection}{\thesection.\Alph{subsubsection}}

\begin{document}
\title{OrbCC: High-Throughput and Low-Latency Data Transport for LEO Satellite Networks}

\author{
\IEEEauthorblockN{Aiden Valentine$^{*}$, Ian Wakeman$^{*\dagger}$ and George Parisis$^{*}$}
\IEEEauthorblockA{School of Engineering and Informatics, University of Sussex$^{*}$\\ Zhejiang Gongshang University$^{\dagger}$\\
\{a.valentine, ianw, g.parisis\}@sussex.ac.uk}
}

\maketitle

\begin{abstract}


The highly dynamic nature of Low-Earth Orbit (LEO) satellite networks introduces challenges that existing transport protocols fail to address, including non-congestive latency variation and loss, transient congestion hotspots, and frequent handovers that cause temporary disconnections and route changes with unknown congestion and delay characteristics. Our contention is that with this increase in complexity, there is insufficient information being returned from the network for existing congestion control algorithms to minimise latency while maintaining high throughput and minimising retransmissions.  Our approach, OrbCC, leverages in-network support to collect per-hop congestion information and uses it to (1) minimise buffer occupancy and end-user latency, (2) maximise application throughput and network utilisation, and (3) rapidly respond to congestion hotspots. We evaluate OrbCC against state-of-the-art transport protocols using OMNeT++/INET-based LEO satellite simulations and targeted micro-benchmarks. The simulations capture RTT dynamics in a LEO constellation, while the micro-benchmarks isolate key characteristics such as non-congestive latency variation and loss, path changes, and congestion hotspots. Results show that OrbCC significantly improves goodput while simultaneously reducing latency and retransmissions compared to existing approaches.
\end{abstract}

\section{Introduction}
\label{sect:introduction}

\begin{figure*}[ht]
  \centering
  \includegraphics[width=0.85\textwidth]{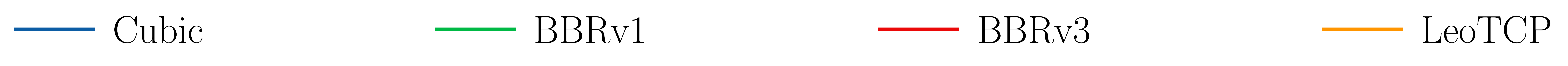}
   \vspace{-0.2cm}
  \begin{subfigure}[b]{0.336\textwidth}
    \centering
    \includegraphics[width=\textwidth]{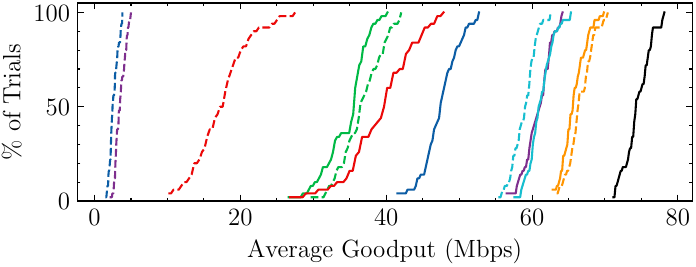}
    \vspace{-0.6cm}
    \caption{CDF of Goodput}
    \label{fig:experiment1Goodput}
  \end{subfigure}\hfill
  \begin{subfigure}[b]{0.32\textwidth}
    \centering
    \includegraphics[width=\textwidth]{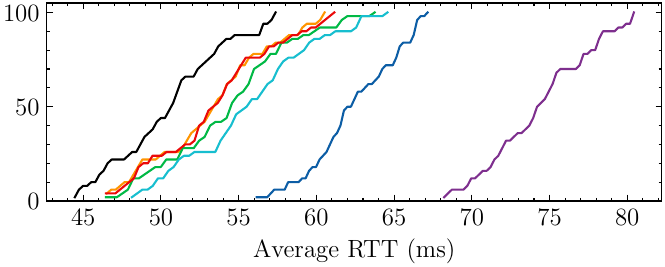}
    \vspace{-0.6cm}
    \caption{CDF of RTT}
    \label{fig:experiment1RTT}
  \end{subfigure}\hfill
  \begin{subfigure}[b]{0.32\textwidth}
    \centering
    \includegraphics[width=\textwidth]{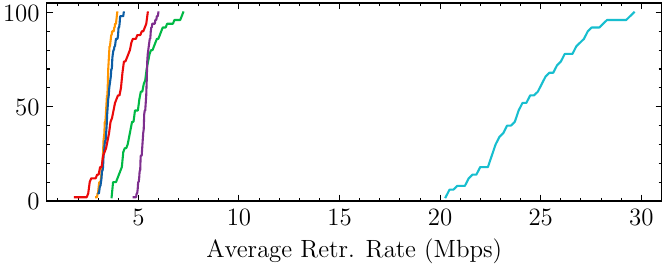}
    \vspace{-0.6cm}
    \caption{CDF of Retransmissions}
    \label{fig:experiment1Retransmissions}
  \end{subfigure}

  \caption{CC Responsiveness: Cumulative Distributions}
  \label{fig:experiment1fig1}
  \vspace{-0.4cm}
\end{figure*}

Low Earth orbit (LEO) satellite networks consist of large constellations of satellites orbiting at altitudes low enough to enable reduced latency compared to traditional geostationary systems. Their orbital parameters are defined by both altitude and inclination. For example, Starlink’s first orbital shell operates at an altitude of 550 km with a 53$^{\circ}$ inclination. Higher inclination angles, closer to 90$^{\circ}$, provide coverage near the poles, while lower inclinations primarily serve equatorial regions. At this altitude, satellites travel at $\sim$7.5km/s, resulting in short visibility windows and necessitating frequent handovers between satellites and ground interfaces \cite{cakaj_parameters_2021, kassing_exploring_2020, cao_satcp_2023}. These networks rely on Ground Stations (GS), which serve as gateways to terrestrial networks, and User Terminals (UT) providing Internet access to end users. GSs are equipped with multiple phased array antennas, enabling simultaneous connections to several satellites \cite{wong_network_2024}. In contrast, UTs are limited to a single-phased array antenna and can maintain only one active satellite link at a time. Early designs focused on Bent Pipe (BP) routing, where ISLs were unavailable and satellites relayed data via GSs. BP paths can offer lower latency over short distances, especially when endpoints are near the same satellite \cite{handley_using_2019}. However, the increasing adoption of inter-satellite links (ISLs) now enables traffic to be routed entirely through space. These ISLs support lower RTTs and higher throughput compared to purely terrestrial paths \cite{hauri_internet_2020}.

LEO satellite networks introduce a set of distinct challenges for data transport protocols and congestion control (CC) mechanisms. Unlike terrestrial networks, LEO constellations operate in a highly dynamic environment where the network topology and routing paths change constantly due to satellite mobility. This results in (1) non-congestive latency variation and loss, (2) transient hotspots and (3) frequent handovers.

\noindent\textbf{Non-congestive latency variation and loss.} Propagation delays fluctuate due to satellite movement, leading to varying distances between UTs, GSs, and satellites. Furthermore, BP routing can amplify this variability, particularly over longer paths \cite{handley_using_2019, valentine_developing_2021}. This complexity is compounded by the structural mesh design of current LEO constellations, where shortest-path routing may not always yield optimal latency and often results in routing through congested or suboptimal links \cite{bhattacherjee_network_2019}. With high bandwidths and large variations of RTTs \cite{kassing_exploring_2020}, buffer resourcing throughout the network must aim for minimal occupancy. In addition, rain fade can cause signal attenuation, resulting in slower data rates, increased latency, or temporary service interruptions \cite{hauri_internet_2020, liu_fade_2009}. 

\noindent\textbf{Handovers.} One of the most critical consequences of satellite mobility in LEO networks is the need for frequent handovers \cite{cao_satcp_2023}. As GSs support multiple simultaneous connections, they are able to conduct soft handovers, where a new satellite link is established before the old one is disconnected. This overlap ensures seamless transitions with minimal risk of packet loss. In contrast, UTs, limited to a single antenna, must perform hard handovers, disconnecting from the current satellite before establishing a new link. This introduces a short connectivity disruption and, potentially, packet loss. In both cases, handovers force traffic to reroute through newly selected paths, which may differ significantly in RTT, available bandwidth, and congestion state.

\noindent\textbf{Transient Hotspots.} Finally, the routing dynamics of LEO networks can cause multiple flows to converge on shared bottlenecks, producing transient congestion hotspots. Prior work has shown that shortest-path routing on LEO mesh topologies can lead to uneven link utilisation \cite{dai_distributed_2021}, particularly across high-demand routes such as transatlantic paths between the US and Europe \cite{gvozdiev_low-latency_2017, kassing_exploring_2020}. These hotspots can lead to buffer buildup, inflated RTTs, and unfair bandwidth distribution among competing flows.

Traditional CC approaches do not consider such dynamic characteristics. To highlight their limitations, we simulate a single flow over a LEO satellite network path with periodic changes in bandwidth, RTT, and random non-congestive loss, updating every 5 seconds to simulate rapid handovers (shown in Figure~\ref{fig:experiment1fig1}).\footnote{The experimental setup and results are discussed further in Section~\ref{sec:responsiveness}.} Loss-based approaches such as Cubic operate with full queues \cite{ha_cubic_2008, barbosa_comparative_2023}, undermining the inherent latency benefits of LEO satellite networks (shown in Figure~\ref{fig:experiment1RTT}). Such approaches also cannot maintain high goodput due to the loss caused by handovers (solid blue line). Under non-congestive loss, Cubic’s performance collapses further (dashed blue line). Other state-of-the-art approaches, such as BBR, aim to operate near the bandwidth-delay product (BDP) by estimating the bottleneck bandwidth and round-trip time (RTT) \cite{cardwell_bbr_2016}. However, BBR relies on periodic probing to update its bandwidth and RTT estimates, which may fail to capture rapid path or latency changes, especially in LEO networks where such changes occur frequently, causing delayed adaptation until the next probe cycle. BBR has also been shown to be unfair when flows with different RTTs compete for the same bottleneck \cite{hock_experimental_2017}. BBRv1, while more resilient to random loss (dashed green line), consistently falls short of fully utilising available bandwidth (solid green line). BBRv3 slightly improves on this (solid red line), but likewise cannot update its model fast enough in the presence of rapid handovers. BBRv3 also suffers greatly with non-congestive loss (dashed red line). These limitations highlight the need for a data transport protocol that can react swiftly to LEO-specific dynamics and sustain high utilisation while keeping queue occupancy low. Recently, bespoke CC schemes tailored to LEO networks have begun to emerge. SaTCP \cite{cao_satcp_2023} is an extension to Cubic, which attempts to predict the occurrence of satellite handovers and route changes to minimise severe reactions to loss. However, it only attempts to mitigate the impact of these disruptive events; when no such event occurs, SaTCP behaves exactly like Cubic, with the issues outlined above (dashed purple line). StarQUIC \cite{kamel_starquic_2024} is a similar approach that freezes the congestion window during handovers, for both BBRv3 and Cubic, and likewise shares the same issues. LeoCC \cite{lai_leocc_2025} is another recent CC algorithm, which extends BBRv1 and attempts to detect path reconfigurations, allowing it to refresh its bottleneck model to avoid relying on outdated bandwidth/RTT samples. While effective in the scenarios it targets, LeoCC incurs repeated startup phases for every reconfiguration, leading to spikes of congestion and loss after each reconfiguration (see Figure \ref{fig:experiment1Retransmissions}). LeoCC additionally relies on ICMP probing every 10ms to detect reconfigurations, with the assumption that there is a separate queue management for ICMP and TCP/UDP traffic. This assumption is validated for Starlink but may not hold for future or non-Starlink deployments.

Our contention is that by increasing the information about network state being returned to senders through in-network telemetry (INT), CC algorithms can ensure low buffer occupancy and high goodput while avoiding unnecessary retransmissions, thereby preserving the performance potential of LEO networks. Motivated by this, we propose OrbCC, which leverages per-hop, switch-reported signals to obtain queue information, allowing the sender to adjust the congestion window according to the current network load. Naively applying existing terrestrial CC designs that employ in-network support \cite{li_hpcc_2019} is insufficient in LEO satellite networks due to the LEO environment violating their key assumptions. OrbCC instead directly tackles the LEO-specific challenges: (1) We can determine the load at the bottleneck and adjust the sending rate to the calculated link capacity. We can minimise buffer occupancy, and, consequently, maintain low latency by adjusting OrbCC's sending rate to be just below the calculated bottleneck BDP. (2) OrbCC can determine and quickly react to path changes, maximising goodput whilst minimising any potential loss. (3) We can quickly react to congestion in the network, reducing the potential for hotspots. (4) To minimise the negative effects of non-congestive loss \cite{cao_satcp_2023, vasisht_l2d2_2021}, and given that OrbCC aims to operate with near-empty buffers at all times, packet loss is not used as an indication of congestion. Building on this, OrbCC introduces three novel mechanisms validated in our evaluation: (i) rapid path-change detection and reaction to sustain goodput across handovers, (ii) improved fairness across diverse path delays, and (iii) an initial-phase strategy that enables new flows to ramp up rapidly and fairly.

\section{OrbCC Design}
\label{sect:design}

OrbCC is built on top of TCP, allowing us to leverage TCP mechanisms such as Selective Acknowledgements (SACK) \cite{floyd_tcp_1996}, timestamps \cite{jacobson_tcp_1992}, and the RACK loss detection algorithm \cite{cheng_rack-tlp_2021}. Switching devices, operating on satellites and GSs, report per-hop network state information, which is appended to the header of each forwarded packet. The use of such in-network support has been explored in the past in the context of data centers \cite{li_hpcc_2019}, but involves major redesigns to ensure efficient and fair operation in LEO satellite networks. Upon the packet's arrival at the receiver, the hop data is extracted and embedded into the corresponding acknowledgement (ACK). With this information, OrbCC senders can assess the bottleneck load on the current path and adapt their sending rates accordingly.




Each switch maintains an average RTT (\texttt{avgRTT}), and the total number of active flows (\texttt{flowCnt}) for every interface. To track the number of flows, each switch identifies flows based on the TCP four-tuple. For the \texttt{avgRTT} value, each switch aggregates the \texttt{baseRTTs} of observed packets using a weighted average that accounts for their \texttt{cwnd} sizes. Specifically, it accumulates two values: the sum of \texttt{baseRTTs} weighted by the inverse of the \texttt{cwnd}, and the sum of squared \texttt{baseRTTs} also weighted by the inverse \texttt{cwnd}. Dividing these two values yields an approximation of the average RTT. These per-interval calculations can be trivially implemented in P4 programmable switches \cite{bosshart_p4_2014, kunze_tofino_2021}. Similarly to XCP \cite{katabi_congestion_2002}, a timer set to the most recently observed average RTT triggers periodic updates of \texttt{flowCnt} and \texttt{avgRTT}. Switches append \texttt{flowCnt} to the header, using the field of the same name, along with the per-hop congestion information. This enables OrbCC senders to calculate their additive increase (AI) factor based on the level of congestion at the bottleneck, as described below. Additionally, switches update the \texttt{pathID} in the header by XOR-ing their own unique \texttt{switchID} with the current \texttt{pathID}. The final \texttt{pathID} is then used by the sender for detecting path changes.  

\begin{figure}[h]
    \centering
    \includegraphics[scale=1.25]{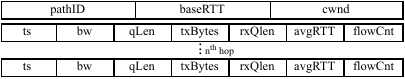}
    \caption{OrbCC Header}
    \vspace{-0.3cm}
    \label{fig:intHeader}
\end{figure}

Figure \ref{fig:intHeader} illustrates the OrbCC header format, which is included in every packet. At transmission, the sender inserts \texttt{pathID}, \texttt{baseRTT}, and \texttt{cwnd}, resulting in approximately 52 bits of fixed per-packet overhead. These fields remain unchanged along the path, except as described above for \texttt{pathID}.


\begin{table}[ht]
\centering
\caption{OrbCC Header Field Descriptions}
\label{tab:intFields}
\renewcommand{\arraystretch}{1.2}
\begin{tabular}{@{}ll@{}}
\toprule
\textbf{Field}      & \textbf{Description} \\ \midrule
\texttt{pathID}     & XOR of per-hop switch IDs to detect path changes \\
\texttt{baseRTT}    & Base RTT estimate from TCP, used in \texttt{avgRTT} estimation \\
\texttt{cwnd}       & Sender’s congestion window, used in \texttt{avgRTT} estimation \\
\texttt{ts}         & Timestamp marking when the packet leaves the queue \\
\texttt{bw}         & Bandwidth of the outgoing link \\
\texttt{qLen}       & Queue length (bytes) when the packet leaves the buffer \\
\texttt{rxQlen}     & Queue length (bytes) when the packet enters the buffer \\
\texttt{txBytes}    & Cumulative bytes transmitted from the buffer \\
\texttt{avgRTT}     & Per-hop average RTT based on observed RTT samples \\
\texttt{flowCnt}    & Number of active flows sharing the bottleneck \\
\bottomrule
\end{tabular}
\vspace{-0.22cm}
\end{table}

The additional fields shown in Figure \ref{fig:intHeader} and described in Table \ref{tab:intFields} represent the information appended by each switch along the path of the packet. The fields \texttt{ts}, \texttt{qLen}, \texttt{rxQlen}, \texttt{txBytes}, and \texttt{bw} provide information about the outgoing switch port and are used to calculate the total utilisation of the outgoing link.  Each switch sets the \texttt{flowCnt} value to its estimate of currently active OrbCC flows using the TCP four-tuple. These per-hop fields contribute approximately 132 bits ($\approx$17 bytes) of overhead. An average hop count among major city pairs is approximately 10 \cite{bhattacherjee_network_2019}, leading to roughly 170 bytes of overhead per packet. This represents an average upper bound, since the packet is largest only near the end of the path. In practice, this overhead is modest relative to MSS-sized packets and is outweighed by OrbCC’s goodput gains from precise congestion feedback. To minimise network overhead, 
receivers only echo back the fields corresponding to the identified current bottleneck switch.






OrbCC senders maintain their \texttt{cwnd} value, using the congestion estimates inferred from the information provided by switches. OrbCC follows an additive increase (AI), multiplicative decrease (MD) approach, growing the window conservatively once every smoothed RTT (sRTT) and reacting quickly to congestion. To avoid bursty traffic, OrbCC employs packet pacing \cite{wei_tcp_2006} at the rate \(\frac{cwnd}{\scalebox{0.95}{$\mathit{s}$}RTT}\). Modern NICs support hardware-based pacing, enabling precise inter-packet spacing with minimal CPU overhead \cite{mittal_timely_2015}.

\noindent\textbf{Per-RTT and Per-ACK Reactions.}
Our approach combines both per-RTT and per-ACK reactions, a strategy shown to enable rapid responses to congestion without triggering overreactions. At the beginning of each sRTT window, the sender sets the previous \texttt{cwnd} as a reference (\texttt{refCwnd}) for both additive and potential multiplicative decreases. The \texttt{cwnd} increases by one AI factor per sRTT. However, if overutilisation is detected, evaluated on a per-ACK basis, OrbCC applies MD, scaled proportionally to the severity of congestion.

\noindent\textbf{Target Utilisation.} 
To estimate the level of bottleneck congestion, the sender computes a utilisation value, $U$, for each hop $i$ along the forward path. A $U$ value of 1 suggests that the link is at full capacity, while greater and lower than 1 implies over- and under-utilisation, respectively. The $U$ value is calculated as follows:

\begin{equation}
\label{eq:utilisation}
U_i = \frac{ack.qLen_i + txRate_i \times avgRTT}{ack.bw_i \times avgRTT}
\end{equation}

This metric uses the hop information to capture the queue occupancy (\texttt{qLen}) and the outgoing rate ($txRate$) as a proportion of the link’s capacity (\texttt{bw}). The $txRate$ is defined as:

\begin{equation}
\label{eq:txRate}
txRate = \frac{ack.txBytes - prevAck.txBytes}{ack.ts - prevAck.ts}
\end{equation}

To focus control on the most congested hop in its path, OrbCC selects the one with the highest $U$ value as the current bottleneck and uses the respective network provided information to adjust its window. To ensure stability and prevent overreaction to transient congestion, OrbCC maintains an exponentially weighted moving average (EWMA) of the selected $U$ value.

\noindent\textbf{Dealing with latency variation.} 
A key challenge in LEO satellite networks is the high variability in propagation delays. When computing the utilisation metric in Equation~\eqref{eq:utilisation}, using a sRTT measurement leads to unfairness across flows with varying latencies. Specifically, flows with higher RTTs will observe lower utilisation values for the same level of congestion, causing them to increase their sending rates more aggressively than lower RTT flows. CC algorithms that rely on feedback signals to manage transmission rates have been shown to become unstable with varying delays \cite{low_dynamics_2002, paganini_scalable_2001}. To address this, OrbCC uses an average RTT in Equation~\eqref{eq:utilisation} which is computed by switches independently, as described above. These values are appended to the headers and echoed back to the sender, ensuring that all flows referencing the same bottleneck compute the same utilisation. 
In OrbCC, the average RTT used in calculating utilisation ensures that window growth is scaled relative to the bottleneck’s conditions, and independently of per-flow propagation delay, making it robust to the varying latencies found in LEO networks.\footnote{In Section \ref{sect:OrbCCDesignEvaluation1}, we experimentally show the dramatic improvement in fairness that this design decision has in OrbCC.}

A key limitation of using the instantaneous per-ACK RTT measurements is that they reflect both propagation and queueing delays. As queues build up along the path, the RTT increases, which in turn inflates the per-hop average RTT computed at switches. This has a compounding effect: the inflated RTT leads to a higher estimated BDP, prompting flows to set their \texttt{cwnd} higher than what is optimal, resulting in greater queue occupancy before overutilisation is detected. To mitigate this, OrbCC leverages the per-interface \texttt{rxQlen} and \texttt{bw} fields, which represent the queue length before the packet is enqueued and the outgoing link bandwidth, respectively. By summing the ratio \(\frac{\texttt{rxQlen}}{\texttt{bw}}\) across all hops, OrbCC estimates the total queueing delay experienced along the path. Subtracting this from the instantaneous per-ACK RTT provides an approximation of the base RTT. To reduce noise, the base RTT is smoothed using an EWMA and appended to each header, enabling switches to compute an average base RTT across flows independent of queueing delays. This decoupling of propagation delay from RTT prevents excessive \texttt{cwnd} growth due to inflated per-ACK RTTs, helping to minimise persistent queue build-up and maintain low latency.

\noindent\textbf{Additive Increase and Multiplicative Decrease.} 
OrbCC employs an AIMD strategy based on the measured link utilisation U. AI is used to grow the \texttt{cwnd}; MD is applied only when utilisation exceeds the target $\eta = 0.95$. This target is chosen to ensure high link utilisation while keeping buffer occupancy low. The AI factor is defined as follows, where \texttt{bw} and \texttt{flowCnt} are the bottleneck's values:

\begin{equation}\label{eq:AI}AI = \frac{bw \times RTT}{flowCnt} \times (1-\eta) \end{equation}

AI is proportional to the estimated BDP of the link, which is computed using the sampled per-ACK RTT rather than the average. This ensures that flows with lower RTTs are not disadvantaged by slower growth. To ensure fair and BDP-proportional growth of flows, the bottleneck bandwidth is divided by the estimate of the number of active flows. This prevents queue build-up as the number of active flows changes. As described above, switches maintain an estimate of the number of active flows, which senders use to compute the AI value. 

When measured utilisation exceeds the target threshold $\eta$, OrbCC applies an MD factor to reduce \texttt{cwnd}. The reduction factor is calculated as the ratio of the current utilisation U to the target $\eta$. This ensures that the reduction is proportional to the degree of overutilisation, meaning larger deviations from the target result in a more aggressive backoff. This design allows OrbCC to maintain high link utilisation while responding swiftly to congestion and emerging hotspots. The AIMD scheme is described as follows, with the \texttt{refCwnd} being updated once every sRTT:


\begin{equation}
cwnd =
\begin{cases}
\frac{refCwnd}{U/\eta} + AI, & \text{U $>=$ $\eta$} \\
refCwnd + AI, & \text{U $<$ $\eta$}
\end{cases}
\end{equation}

\noindent\textbf{Initial Phase.} When new flows join a path already shared by many existing flows, they must be able to ramp up their sending rate quickly enough, whilst not saturating bottleneck buffers or hurting the performance of existing network flows. Since the AI factor is inversely proportional to the estimated number of flows, newly started flows may begin ramping up their sending rate too conservatively, delaying convergence to their fair share of bandwidth. To address this, OrbCC introduces a bespoke start-up mechanism. When a new flow is created, it starts the initial phase. The flow then sends a small number of packets into the network to collect hop feedback. Once it receives the first header, the flow waits for an \texttt{avgRTT} (taken from the respective header) before increasing its \texttt{cwnd}. This delay allows the switch-estimated flow count to reflect both the current flow, and, potentially, other newly arrived flows. After this waiting period, the sender computes its AI as 95\% of the bottleneck BDP divided by the estimated flow count, rather than 5\% in Algorithm \ref{eq:AI}. Using 95\% as the AI allows newly arriving flows to converge rapidly to their fair rate while avoiding the need for unnecessary multiplicative decreases once the initial phase ends.

\noindent\textbf{Reacting to path changes.} To maintain stability when path changes and handovers occur, OrbCC is designed to respond rapidly to RTT and congestion changes, without overreacting to packet loss events. Unlike  loss-based approaches \cite{ha_cubic_2008, floyd_newreno_2004}, which reduce the \texttt{cwnd} after three duplicate ACKs, OrbCC adjusts its window solely based on the provided in-network information. This enables the sender to distinguish between congestive and non-congestive loss events. For retransmission and loss recovery, OrbCC employs the TCP SACK extension \cite{floyd_tcp_1996} together with RACK \cite{cheng_rack-tlp_2021}. SACK enables precise identification of lost packets, while RACK uses timers instead of packet ordering to detect losses, allowing OrbCC to be more resilient to reordering. To mitigate the impact of reordering due to path changes, OrbCC leverages the timestamp fields embedded in its headers to ensure that only current network provided information is used for congestion estimation. Additionally, OrbCC uses the \texttt{pathID} field to detect path changes, allowing it to prioritise ACKs associated with the new path while safely ignoring outdated ACKs from old routes. To detect path changes, senders compare the XORed \texttt{pathID} from incoming ACKs to their currently stored value. If a new path is detected, the sender updates its stored \texttt{pathID}, but only once per sRTT interval to avoid reacting to reordering. If no change is detected, the existing value is maintained.

\section{Experimental Evaluation}
\label{sect:evaluation}
We evaluate OrbCC using packet-level simulations in OMNeT++ \cite{noauthor_omnet_nodate} and INET \cite{noauthor_inet_nodate}, extended with a LEO satellite model\footnote{Current LEO simulation models\cite{valentine_developing_2021, kassing_exploring_2020} do not support hard handovers at the user terminal. Therefore, we design separate experiments in Sections~\ref{sec:responsiveness} and $h$ to specifically evaluate these scenarios.} \cite{valentine_developing_2021} supporting BP and ISL topologies. Our setup simulates Starlink’s first shell with 100 ground stations, updating routing every 100ms to reflect dynamic connectivity and RTT variation. Buffer sizes match the BDP of the longer RTT path. We implemented Cubic, BBRv1, BBRv3, SaTCP, LeoCC and OrbCC refining INET’s SACK implementation to better reflect the Linux kernel counterpart and added support for the RACK loss recovery mechanism \cite{cheng_rack-tlp_2021}. RACK reflects common practice in modern TCP variants. Source code is available.\footnote{https://figshare.com/s/334dbee6891e4edf6f9e} We validated our models against their Linux kernel counterparts within an emulation-based testbed, showing agreement in performance trends and protocol behaviour. SaTCP and LeoCC both modify CC behaviour in response to handovers; outside of these events it behaves like its underlying CC. We therefore include SaTCP and LeoCC only in experiments that explicitly model hard handovers. We note that the original SaTCP design assumes user-terminals provide live information on its handovers. In our simulations, handover events are known and provided to the protocol. Unless stated, links are 100Mbps and results average five runs.


Using the LEO constellation model \cite{valentine_developing_2021}, we first examine fairness and delay when two flows share the same path, and then when flows traverse different RTT paths that intermittently converge on a bottleneck. Since the model does not capture hard handovers, we evaluate Cubic, BBRv1, BBRv3, and OrbCC in these experiments.


\begin{figure*}[!ht]
    \centering
    \hspace*{-34mm}
    \begin{subfigure}[t]{0.09\textwidth}
        \vspace*{-38.2mm}
        \hspace*{0.05cm}
        \centering
        \includegraphics[width=\linewidth,clip,trim=0 0 0 0]{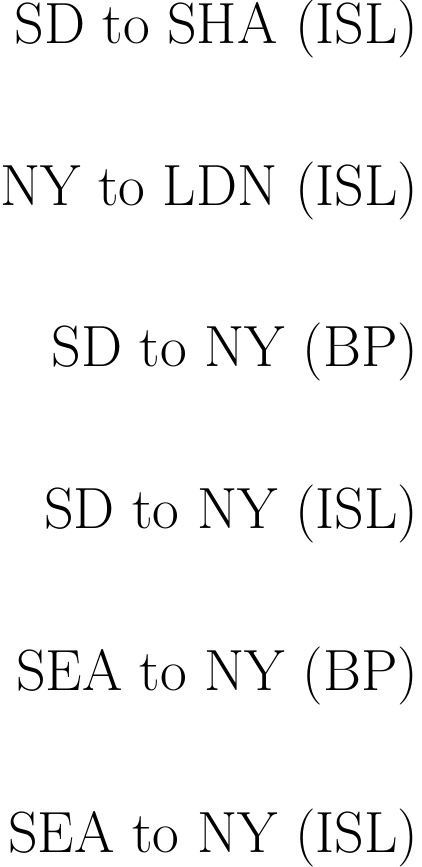}
    \end{subfigure}
    \begin{subfigure}[t]{0.285\textwidth}
        \includegraphics[width=\linewidth]{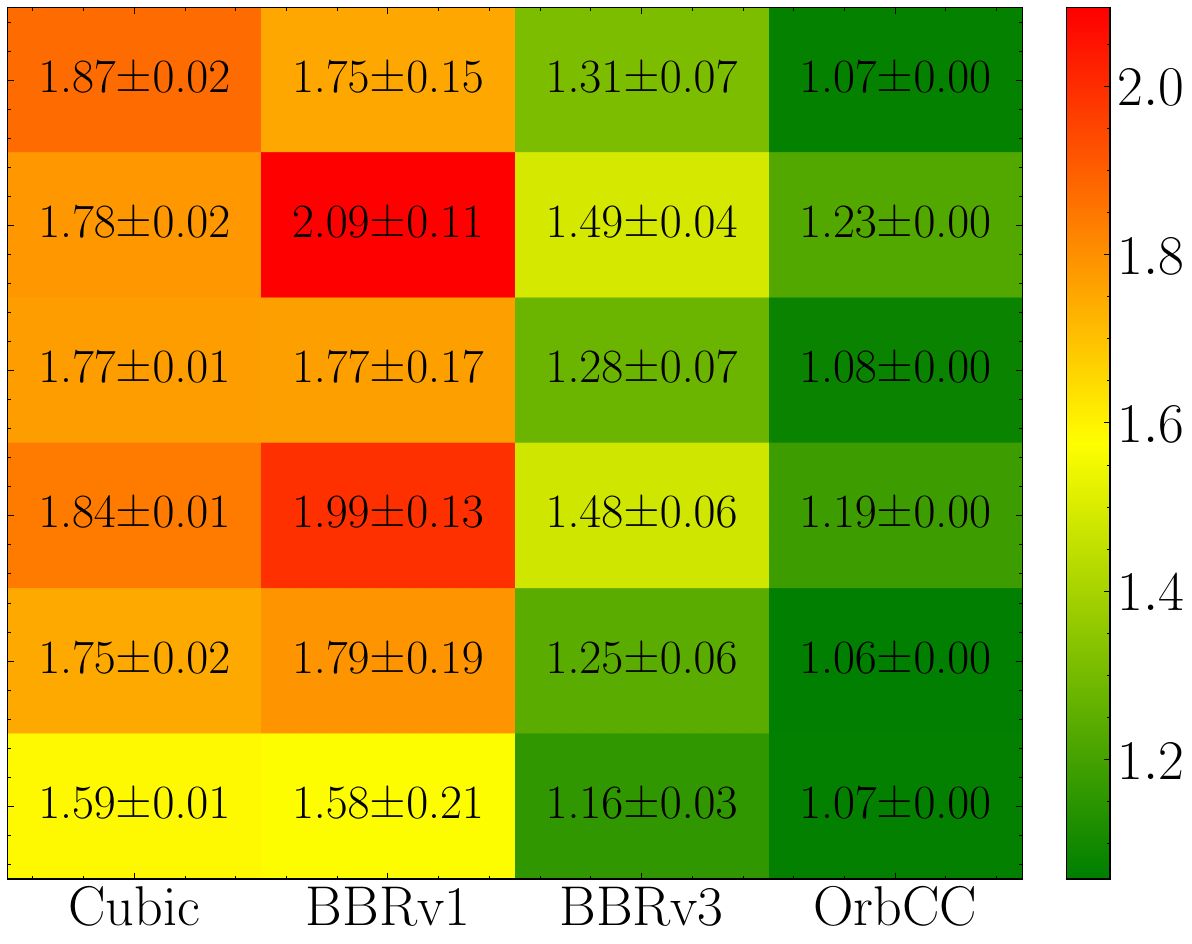}
        \caption{Average normalised delay}
        \label{fig:leoExperiment1}
    \end{subfigure}
    \begin{subfigure}[t]{0.323\textwidth}
        \includegraphics[width=\linewidth]{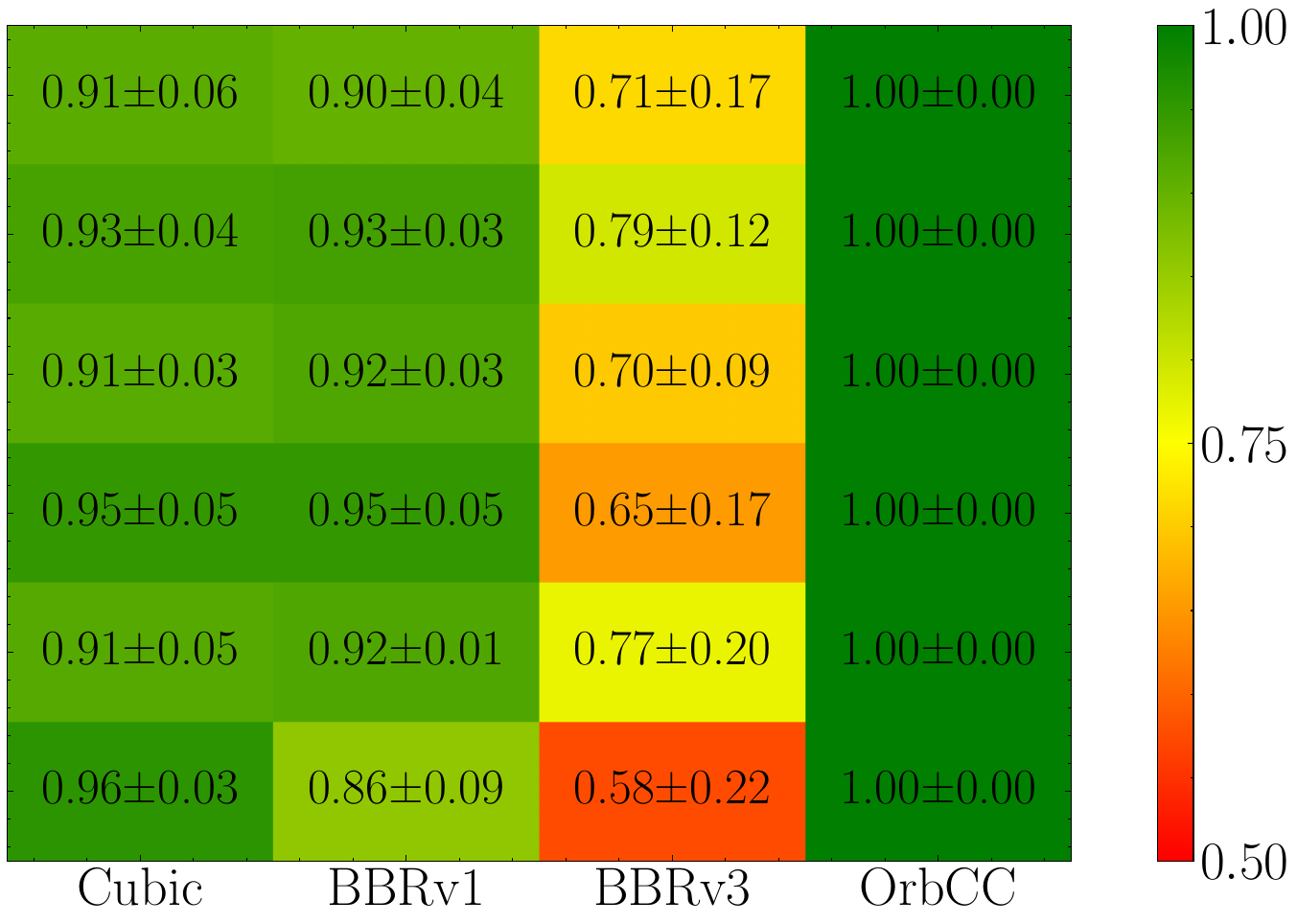}
        \caption{Average goodput ratio}
        \label{fig:leoExperiment2}
    \end{subfigure}
    \begin{subfigure}[t]{0.12\textwidth}            
        \centering
        \vspace*{-40mm}                            
        {\scriptsize
        \renewcommand{\arraystretch}{0.8}%
        \setlength{\tabcolsep}{2pt}%
        \begin{tabular}{lcccc}
        \toprule
        \textbf{Pair} & \textbf{Cubic} & \textbf{BBRv1} & \textbf{BBRv3} & \textbf{OrbCC} \\
        \midrule
        \multicolumn{5}{c}{\textbf{Average Goodput Ratio}} \\
        \midrule
        Pair1 & \cellcolor[rgb]{0.83,0.91,0.80}0.96±0.02 & \cellcolor[rgb]{1.00,0.96,0.80}0.71±0.01 & \cellcolor[rgb]{0.97,0.98,0.80}0.79±0.05 & \cellcolor[rgb]{0.80,0.90,0.80}0.99±0.01 \\
        Pair2 & \cellcolor[rgb]{0.83,0.92,0.80}0.96±0.03 & \cellcolor[rgb]{1.00,0.80,0.80}0.40±0.01 & \cellcolor[rgb]{1.00,0.99,0.80}0.74±0.16 & \cellcolor[rgb]{0.82,0.91,0.80}0.97±0.00 \\
        \midrule
        \multicolumn{5}{c}{\textbf{Average Goodput (Mbps)}} \\
        \midrule
    Pair1 & 75.79 & 76.28 & 76.41 & 78.46 \\
    Pair2 & 61.09 & 59.38 & 60.74 & 61.85 \\
    \midrule
        \multicolumn{5}{c}{\textbf{Average Normalised Delay}} \\
        \midrule
        Pair1 &
        \cellcolor[rgb]{0.90,0.94,0.80} 1.38$\pm$0.01 &
        \cellcolor[rgb]{0.85,0.92,0.80} 0.95$\pm$0.01 &
        \cellcolor[rgb]{0.82,0.91,0.80} 0.81$\pm$0.02 &
        \cellcolor[rgb]{0.80,0.90,0.80} 0.73$\pm$0.00 \\
        
        Pair2 &
        \cellcolor[rgb]{1.00,0.80,0.80} 2.90$\pm$0.05 &
        \cellcolor[rgb]{1.00,0.91,0.80} 2.27$\pm$0.01 &
        \cellcolor[rgb]{1.00,0.99,0.80} 1.86$\pm$0.07 &
        \cellcolor[rgb]{0.90,0.94,0.80} 1.44$\pm$0.00 \\
        \bottomrule
        \end{tabular}
        }
        \vspace*{2.5mm} 
        \captionsetup{width=50mm} 
        \caption{Average goodput ratio, goodput and normalised delay}
        \label{fig:leoExperiment3}
    \end{subfigure}
    \caption{Experimentation with the LEO simulation model: (a) and (b) two flows on the same path; and (c) two flows on different RTT paths that intermittently converge at shared bottlenecks.}
    \vspace{-0.3cm}
    \label{fig:leoExperiments}
\end{figure*}


\paragraph{\textbf{Path Sharing - Same RTTs}}
\label{sect:leoModel1}
Figures~\ref{fig:leoExperiment1} and \ref{fig:leoExperiment2} show the average normalised delay and goodput ratio for two concurrent flows sharing a LEO satellite network path. Cubic flows get a fair share of bandwidth. This comes at the cost of significant delay inflation (up to 1.87$\times$) due to Cubic’s buffer filling behaviour (Figure~\ref{fig:leoExperiment1}). BBRv1 maintains good fairness; it tends to overestimate the available bandwidth when multiple flows share a bottleneck \cite{scholz_towards_2018}. This results in persistent queue build-up and high delay inflation of up to 2.09$\times$. Unlike BBRv1, BBRv3 achieves lower delay inflation by incorporating loss as a signal to adjust its pacing rate, and by probing bandwidth less frequently. This loss sensitivity can lead to aggressive rate reductions for some flows, resulting in poor fairness. When the base RTT increases, BBRv3 continues pacing at an outdated BDP, leading to temporary under-utilisation, and potentially unfairness, until its 10-second RTT filter expires, and a new probe refreshes its estimates. While BBRv1 shares this mechanism, it performs better by filling buffers more aggressively. OrbCC achieves an excellent goodput ratio between flows and minimal delay inflation due to its use of hop-level data, allowing it to react quickly to path and latency changes. OrbCC provides fair bandwidth sharing even in dynamic conditions by using the estimated number of flows at the bottleneck to compute a proportional additive increase for each flow.

\paragraph{\textbf{Intermittent Path Sharing - Different RTTs}}
\label{sect:leoModel2}

Figure~\ref{fig:leoExperiment3} presents the average goodput ratio for two flows on different paths that intermittently share a bottleneck. The experiment includes two pairs of paths: Pair 1, with paths between New York to London and New York to St. John’s, Canada, which share a bottleneck for approximately 40\% of the simulation; and Pair 2, with paths between San Diego to New York and Lawrence, Kansas to New York, with about 70\% bottleneck overlap. Cubic performs well in both pairs, despite its tendency to favour lower RTT flows that react more quickly to loss \cite{giacomoni_reinforcement_2024}. This is because the results reflect average goodput across the full simulation that includes long periods where flows do not share a bottleneck; this reduces the impact of the short-term unfairness on the goodput ratio. BBRv1 shows substantial unfairness, particularly in Pair 2, as the bottlenecks are shared for a longer period of time. This stems from BBRv1’s bandwidth estimation and pacing, which favours flows with larger in-flight data. BBRv3 slightly improves fairness over BBRv1 through its updated probe down phase, where the pacing gain increases from 0.75 to 0.91, causing the flow to reduce its rate less aggressively and limiting the ability of lower RTT flows to dominate the bandwidth. The normalised delay for all protocols mirrors characteristics from the previous experiment, with OrbCC achieving both lower delays and excellent fairness by ensuring that all flows compute a unified utilisation value through its \texttt{avgRTT} mechanism when competing for bandwidth.


\paragraph{\textbf{Responsiveness}}
\label{sec:responsiveness}
We revisit the motivating experiment from the Introduction: a single flow over a LEO path with bandwidth and RTT changing every 5s, where each reconfiguration induces a hard handover outage of 45-120ms. Bandwidth varies from 50–100Mbps and RTT from 1–100ms, with buffers set to the mean BDP. Each protocol is run 50 times.


\begin{figure}
    \centering
    \includegraphics[scale=0.8]{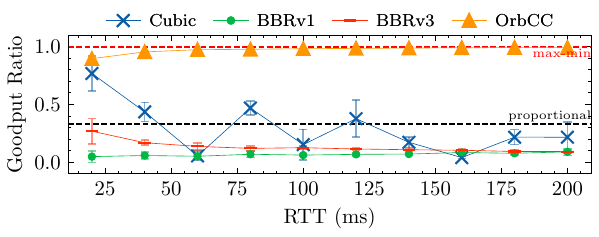}
    \vspace{-0.2cm}
    \caption{Goodput ratio between the flow experiencing multiple bottlenecks and the highest goodput achieved by any single-bottleneck flow, both experiencing the same RTT (x-axis).}
    \label{fig:parkinglot}
    \vspace{-0.3cm}
\end{figure}

As shown in Figure \ref{fig:experiment1fig1}, both BBRv1 (solid green line) and BBRv3 (solid red line) consistently underutilise the available bandwidth due to their reliance on a minimum RTT filter, causing slow reaction to path changes. Cubic adapts quickly to bandwidth changes through its cubic window growth function (solid blue line), but this comes at the cost of significant latency inflation due to buffer overfilling. Handover-induced losses repeatedly trigger Cubic's multiplicative backoff, leading to sustained underutilisation. SaTCP, by freezing its congestion window across reconfigurations, avoids this handover-driven underutilisation, but still incurs large delay due to buffer build-up (solid purple line). LeoCC attains similarly high goodput due to it resetting its model and entering startup post-reconfiguration (solid teal line). However, LeoCC's startup exponentially ramps-up until a stable bandwidth is observed for three consecutive RTTs, inflating queues and inducing loss. In contrast, OrbCC reacts to reconfigurations using its per-hop congestion information, tracking bandwidth changes accurately with minimal delay inflation (solid orange line). Its explicit 95\% utilisation target prevents persistent queue build-up and ensures consistently low latencies.

In the presence of non-congestive loss, both Cubic and SaTCP’s goodput degrades dramatically because both CCs interpret loss as a congestion signal, unnecessarily reducing its sending rate (dashed blue and purple line in Figure \ref{fig:experiment1Goodput}). Both BBRv1 and its derivative, LeoCC, ignore loss and therefore avoid overreacting, performing similarly to the no-loss cases (dashed green and teal lines). BBRv3 continuously probes for bandwidth unless loss exceeds its 2\% threshold, however, it still reacts aggressively even to mild loss, leading to poor goodput performance (dashed red line). OrbCC remains robust in lossy environments as it does not use loss as a congestion signal.


\paragraph{\textbf{Multi-bottleneck fairness}} 
\label{sec:parking_lot}

In LEO satellite networks, the mesh-like connectivity and frequent routing changes could cause flows to traverse multiple congested links \cite{gvozdiev_low-latency_2017}. To evaluate fairness under these conditions, we simulate  several shared bottlenecks. We used a parking lot topology, with one flow crossing three bottlenecks and three flows crossing one of those bottlenecks each. We evaluate performance through the prism of max-min fairness, which ensures that no flow can increase its rate without reducing that of a flow with equal or lesser throughput \cite{crowcroft_differentiated_1998}, and proportional fairness, which seeks to maximise overall efficiency while allowing unequal sharing when it improves aggregate utility. We simulate 0.2$\times$, 1$\times$, and 4$\times$ BDP queues. We present results only for the 1$\times$ BDP case for clarity; performance was very similar in the other setups.

Figure~\ref{fig:parkinglot} shows the goodput ratios of the multiple bottleneck setup. Cubic demonstrates behaviour close to proportional fairness, which becomes more pronounced as buffer sizes increase. Since Cubic reduces its window in response to packet loss, the flow crossing multiple bottlenecks encounters more frequent reductions, as it is more likely to encounter loss across several points in the path. BBRv3 and BBRv1 both exhibit fairness closer to proportional fairness. Although BBRv1 and BBRv3 use different probing mechanisms, in this homogeneous flow setup, flows traversing a single bottleneck end up probing roughly three times more frequently than the multi-bottlenecked flow, contributing to a fairness pattern closer to proportional fairness. OrbCC, on the other hand, maintains strong fairness in the presence of multiple bottlenecks. By explicitly tracking the number of active flows and computing per-hop utilisation estimates, OrbCC enables balanced bandwidth sharing across all flows, leading to bandwidth allocations that approximate max-min fairness.

\paragraph{\textbf{Inter-RTT Fairness}}
\label{sect:interRTTfairness}

\begin{figure}
    \centering
    \includegraphics[scale=0.8]{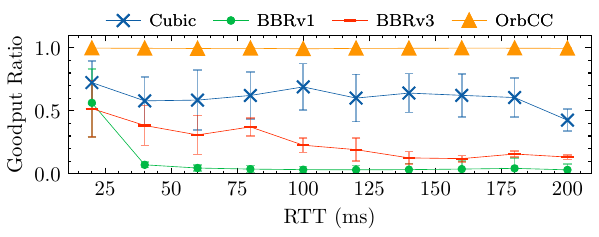}
    \vspace{-0.2cm}
    \caption{Goodput ratio of two competing flows. Starting flow has 20ms RTT and joining flows' RTT is shown on the x-axis.}
    \label{fig:interRTT}
    \vspace{-0.3cm}
\end{figure}

We next examine inter-RTT fairness, where flows competing for the same bottleneck experience different RTTs. One flow has a fixed RTT of 20ms, while the second flow's RTT is varied. The queue size is set to $1\times$ the BDP of the higher RTT flow. The first flow is active for 2000 RTTs. The second flow starts 500 RTTs after the first flow. To allow CC schemes time to converge, we measure goodput over the final 500 RTTs of the experiment. Figure~\ref{fig:interRTT} presents the resulting goodput ratios. Cubic’s RTT bias resurfaces strongly here, as the flows with lower RTTs dominate. The BBR variants experience unfairness as RTTs diverge. This is because the flow experiencing a higher RTT will push more bytes in flight as a result of measuring a higher BDP, subsequently claiming higher buffer occupancy, leading to unfairness \cite{scholz_towards_2018}. BBRv3 offers slightly improved fairness, as its probe down phase dampens the aggressiveness of rate reductions, but it still does not fully resolve its RTT unfairness. OrbCC maintains consistent fairness, with goodput ratios close to 1 regardless of RTT disparity. This mirrors its earlier performance using the LEO simulation model, where its consistent utilisation calculation enabled fair bandwidth sharing. 


\begin{figure}
    \centering
    \includegraphics[scale=0.8]{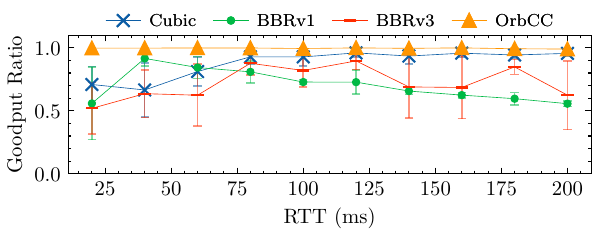}
    \vspace{-0.2cm}
    \caption{Goodput ratio of two competing flows both experiencing the same RTT (shown on x-axis).}
    \label{fig:intraRTT}
    \vspace{-0.3cm}
\end{figure}


\paragraph{\textbf{Intra-RTT Fairness}}
\label{sect:intraRTTfairness}
We now assess intra-RTT fairness, where two flows with the same base RTT compete over a shared bottleneck, otherwise repeating the methodology described in the inter-RTT experiment. In Figure \ref{fig:intraRTT}, the base RTT of the flows is varied, as shown on the x-axis. Cubic demonstrates strong fairness across all RTT values, similar to its behaviour in Section~\ref{sect:leoModel1}. Both BBRv1 and BBRv3 perform slightly worse than Cubic. Upon closely inspecting individual runs, we observe that flows of both protocols take a long time to converge after the second flow is started, in comparison to OrbCC and Cubic. OrbCC consistently achieves the best fairness, maintaining near-optimal goodput ratios across all tested base RTT values.

Soft and hard handovers can cause traffic shifts, leading to packet loss and transient congestion. We conduct one simulation of soft, lossless handovers, and the other looks further into hard handovers following Section \ref{sec:responsiveness}, evaluating how increasing the number of concurrent flows affects aggregate goodput and delay under hard handover disruptions.

\paragraph{\textbf{Soft Handovers}}
\label{sect:softHandovers}

We simulate a soft handover scenario between a GS and two satellites along a LEO satellite network path. Ground station handovers typically enable near-seamless transitions with minimal packet loss \cite{cao_satcp_2023}, although packet reordering may still occur. To evaluate OrbCC’s responsiveness to such events, two flows are initially routed over a 20ms RTT path. At 100 seconds, these flows are redirected to a different network path that already serves two flows with a higher RTT (50ms). At 200 seconds, the rerouted flows return to their original lower RTT path. 
To assess the impact of redirection, we measure the average normalised goodput and Jain’s fairness index after the path change (``Change 1''), and again after the flows return to their original paths (``Change 2''). Measurements are taken over the first 100 RTTs after each change to evaluate the protocols' ability to adapt quickly and fairly.


\begin{figure}
    \centering
    \includegraphics[scale=0.8]{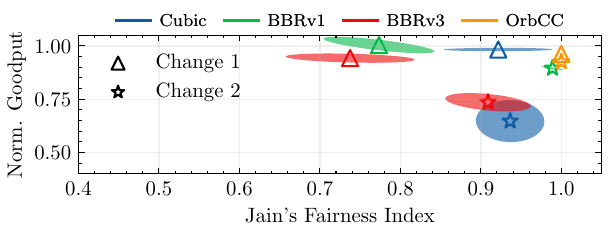}
    \vspace{-0.2cm}
    \caption{Aggregate normalised network goodput and normalised delay for 4 competing flows.}
    \label{fig:crosspath}
    \vspace{-0.3cm}
\end{figure}

Cubic and the BBR variants show higher unfairness during the first change (triangles in Figure \ref{fig:crosspath}), consistent with the inter-RTT unfairness observed above. Following the second change at 200 seconds (Change 2), the experiment transitions into an intra-RTT setup, with both flow groups operating under their respective base RTTs, showing once again similar fairness to the intra-RTT experiment. OrbCC maintains high fairness during both transitions. The high fairness during Change 1 is attributed to its \texttt{avgRTT} and switch-reported path signals. In terms of bandwidth exploration, Cubic does not capture much of the available bandwidth in Change 2, (blue star in Figure \ref{fig:crosspath}). This is attributed to its concave window growth period, which must elapse before Cubic can explore beyond its previously seen maximum congestion window. A higher queue size would allow Cubic to carry a much larger congestion window into Change 2, resulting in a faster bandwidth exploration, at the cost of a higher RTT. The BBR variants must enter the probe bandwidth phase to begin bandwidth exploration. BBRv1 is much quicker compared to BBRv3, as the former is RTT based and the latter is wall-clock-based. OrbCC also maintains high goodput during both changes due to it allowing senders to immediately determine the characteristics of the new path once the first ACK is received after the change. This is further enabled by its use of the \texttt{pathID} field, which allows the sender to accurately detect the path change and immediately prioritise feedback from the new route.

Averaged over 100 RTTs following the first path change, Cubic’s RTT reaches 62.7ms for the rerouted flow on the 20ms base RTT path and 94.57ms for the constant flow on the 50ms base RTT path, indicating substantial queue buildup at the bottleneck. BBRv1 shows similar inflation, with RTTs of 63.45ms (20ms path) and 92.44ms (50ms path), consistent with earlier results in the LEO simulation model (Section~\ref{sect:leoModel2}), where BBRv1 tends to overestimate available bandwidth under multi-flow conditions. BBRv3 performs slightly better, recording 43.63ms (20ms path) and 78.74ms (50ms path) RTTs, due to it reacting to loss, as well as probing for bandwidth less frequently than BBRv1. OrbCC exhibits the lowest delay inflation of all protocols, with RTTs of just 21.50ms (20ms path) and 51.50ms (50ms path). These closely track their respective base RTTs, as OrbCC explicitly targets 95\% utilisation, keeping queueing minimal even during path changes.

\begin{figure}
    \centering
    \includegraphics[scale=0.65]{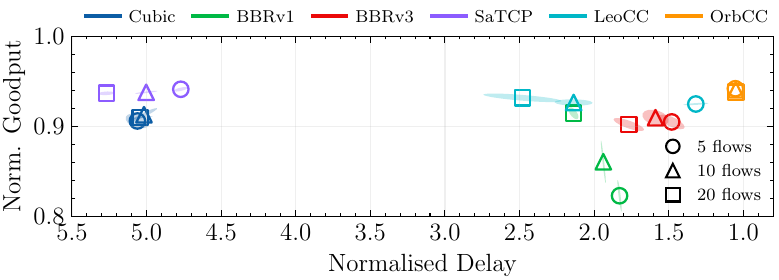}
    \vspace{-0.2cm}
    \caption{Normalised goodput vs normalised delay of multiple flows sharing the bottleneck}
    \label{fig:exp11}
    \vspace{-0.3cm}
\end{figure}

\paragraph{\textbf{Hard Handover Efficiency}}
\label{sect:hardHandovers}
We next evaluate efficiency of multiple flows sharing a bottleneck, in the presence of periodic hard handovers. We schedule 5, 10, and 20 flows on a 50ms path (buffer set to 5$\times$ BDP), staggering flow start times within the first 100s and measuring aggregate goodput and mean delay inflation over the 100-150s interval where flows co-exist. Hard handovers occur every 15s to match the rate at which Starlink UTs perform link reconfigurations \cite{starlink_performance}. Figure~\ref{fig:exp11} shows that Cubic and SaTCP sustain high goodput but do so with the largest delay inflation across all flow counts, reflecting persistent queue build-up. Both BBR versions achieve similarly high utilisation, yet their delay remains clustered around 2 inflation, consistent with BBR’s tendency to maintain an elevated in-flight target and induce queuing under competition \cite{bbrv3wired}. LeoCC attains high goodput as well, but exhibits greater delay, consistent with repeated startup phases after reconfiguration. Finally, OrbCC combines high goodput with minimal delay. By relying on information reported by switches and not using packet loss as a congestion signal, it maintains accurate congestion estimates and resumes sending at the previously established rate once the path is reconfigured.


We now evaluate two key design components of OrbCC that enable high performance in LEO satellite networks: (1) the use of a per-switch \texttt{avgRTT} to ensure fair utilisation across flows with varying propagation delays, and (2) the initial phase mechanism, which enables faster convergence during flow start-up, particularly when many flows contend for the same bottleneck.

\begin{figure}[h]
    \centering
    \vspace{-0.15cm}
    \includegraphics[scale=0.8]{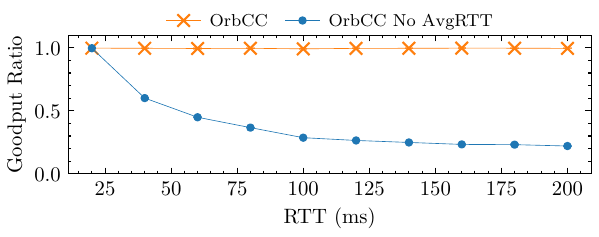}
    \vspace{-0.2cm}
    \caption{Goodput ratio of two competing flows. Starting flow has 20 ms RTT and the joining flow’s RTT is shown on x-axis.}
    \vspace{-0.3cm}
    \label{fig:avgRTTGoodputRatio}
\end{figure}

\begin{figure}[h]
    \centering
    \vspace{-0.15cm}
    \includegraphics[scale=0.8]{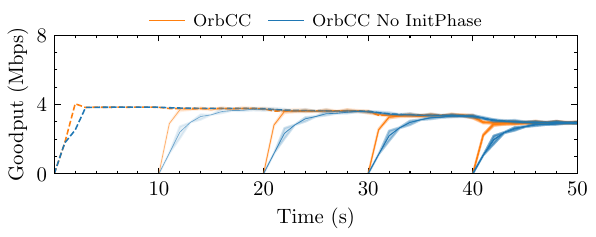}
    \vspace{-0.2cm}
    \caption{Goodput evolution for the initial 50 aggregated flows (dashed lines) and 15 late starting flows (solid lines).}
    \label{fig:intPhaseGoodputEvolution}
    \vspace{-0.3cm}
\end{figure}

\paragraph{\textbf{OrbCC Average RTT}}
\label{sect:OrbCCDesignEvaluation1}
Figure~\ref{fig:avgRTTGoodputRatio} evaluates the effect of OrbCC’s \texttt{avgRTT} mechanism using the same setup as in Section~\ref{sec:responsiveness}, where two competing flows share a bottleneck but experience varying RTTs. We experiment with buffer sizes set to 1$\times$ the BDP of the higher RTT flow. The baseline protocol uses each flow’s sampled per-ACK RTT directly in its utilisation computation, while OrbCC applies a per-hop, switch-reported average RTT. Without \texttt{avgRTT}, flows with higher RTTs observe lower utilisation values for the same level of congestion, increasing their sending rates more aggressively, resulting in significant unfairness. OrbCC’s average RTT design ensures that both flows compute a consistent utilisation value at each hop, leading to fair bandwidth sharing regardless of RTT.

\paragraph{\textbf{OrbCC Initial Phase}}
\label{sect:OrbCCDesignEvaluation2}
We now evaluate the effectiveness of OrbCC’s initial phase mechanism in enabling fast convergence for new flows in a shared bottleneck scenario. The experiment uses a dumbbell topology, where all flows traverse the same bottleneck link with a fixed RTT of 100ms and a bandwidth of 200Mbps. Initially, 50 flows start within a 5 RTT window. Additional flows then join incrementally: 1 new flow at 10s, 2 more at 20s, 4 at 30s, and 8 at 40s. Figure~\ref{fig:intPhaseGoodputEvolution} shows the evolution of goodput over time for all flows. This setup tests whether OrbCC’s initial phase design allows newly arriving flows to quickly ramp up to their fair share without affecting the goodput of existing flows.

As shown in Figure~\ref{fig:intPhaseGoodputEvolution}, without the initial phase, new flows take a large amount of RTTs to reach their fair share of bandwidth. With the initial phase enabled, OrbCC waits one \texttt{avgRTT} after receiving in-network information. This allows the sender to obtain an accurate estimate of the number of competing flows before increasing its rate. As a result, newly arriving flows ramp up smoothly and quickly, without disrupting existing traffic. Despite being more aggressive during start-up, utilising the initial phase introduces only minimal additional queuing. The average RTT of the first 50 flows, measured over 100 RTTs at the start of the simulation, is 104.68ms with the initial phase enabled, compared to 104.13ms without. When eight additional flows join at 40 seconds OrbCC has an average RTT of 105.48ms RTT with the initial phase, versus 104.18ms without. These results confirm that OrbCC can achieve rapid convergence at startup without significantly compromising latency, as both new and existing flows can quickly adapt to changes in network conditions.

\section{Discussion and Conclusion}
\label{sect:conclusion}

OrbCC targets operator-managed LEO constellations under a single administrative domain \cite{ma_network_2023}, enabling uniform deployment across satellites and user terminals; we therefore designed fairness among competing OrbCC flows. This deployment model is consistent with other LEO approaches \cite{cao_satcp_2023, lai_leocc_2025} that assume gateway/middlebox assistance, making TCP-friendliness within the satellite segment a secondary concern. In this setting, performance-enhancing proxies (PEPs) can also be practical \cite{yuan2025internet}: gateways can split or proxy connections at the constellation edge, isolating terrestrial endpoints from LEO-specific dynamics and allowing OrbCC to operate optimally within the satellite segment while preserving end-to-end semantics where required. Compatibility with conventional end-to-end CC is primarily needed at terrestrial interconnects, where gateways can hand off or translate between OrbCC within the constellation and an end-host CC approach outside it.

The current design of OrbCC assumes that its in-network signalling mechanisms are available at all intermediate routers along the path. Under a partial or incremental deployment of OrbCC switches, packets would carry incomplete network state information, potentially limiting the sender’s ability to accurately locate bottlenecks and distinguish between propagation and queueing delays. We leave the design and evaluation of mechanisms for robust operation in such environments as future work, but foresee that OrbCC could fall back to utilising alternate state-of-the-art LEO CC approaches \cite{lai_leocc_2025}.

Through a combination of micro-benchmarks and LEO constellation simulations using OMNeT++/INET, we demonstrated that OrbCC, in LEO satellite networks, achieves high goodput while significantly reducing RTT and retransmissions compared to state-of-the-art schemes such as Cubic and BBR, as well as recent LEO-specific approaches such as LeoCC. Our experiments highlight OrbCC’s resilience to dynamic path changes, soft and hard handovers, and large RTT variations at bottlenecks, all common in LEO deployments.

\bibliographystyle{IEEEtran}
\bibliography{references.bib}

\end{document}